\documentclass[reprint,aps,pre]{revtex4-1}
\usepackage{graphicx}
\usepackage{bm}

\newcommand{\V}[1]{\bm{#1} } 

\newcommand{\Ave}[1]{\left\langle {#1} \right\rangle} 

\newcommand{\lb}{\left(}
\newcommand{\rb}{\right)}
\newcommand{\lbb}{\left\{}
\newcommand{\rbb}{\right\}}

\newcommand{\T}[1]{\tilde{#1}}
\newcommand{\Req}[1]{eq.\ (\ref{eq:#1})}

\newcommand{\Rfig}[1]{Fig.\ \ref{fig:#1}}

\newcommand{\Rfigss}[2]{Figs.\ \ref{fig:#1}-\ref{fig:#2}}

\newcommand{\Lfig}[1]{\label{fig:#1}}
\newcommand{\Leq}[1]{\label{eq:#1}}

\newcommand{\be}{\begin{eqnarray}}
\newcommand{\ee}{\end{eqnarray}}
\newcommand{\ba}{\begin{array}}
\newcommand{\ea}{\end{array}}
\newcommand{\no}{\nonumber}

\newcommand{\subbe}{\begin{subequations}}
\newcommand{\subee}{\end{subequations}}
\newcommand{\argmax}{\mathop{\rm arg~max}\limits}

\begin{document} 

\preprint{APS/123-QED}

\title{Multiple peaks of species abundance distributions induced by sparse interactions}
\author{Tomoyuki Obuchi$^{1}$}
\email{obuchi@c.titech.ac.jp}
\author{Yoshiyuki Kabashima$^{1}$}
\author{Kei Tokita$^{2}$}\affiliation{$^{1}$
Department of Mathematical and Computing Science, Tokyo Institute of Technology, Yokohama 226-8502, Japan
\\
$^{2}$Graduate School of Information Science, Nagoya University, Nagoya 464-8601, Japan
}

\date{\today}

\begin{abstract}
We investigate the replicator dynamics with ``sparse'' symmetric interactions which represent specialist-specialist interactions in ecological communities. By considering a large self interaction $u$, we conduct a perturbative expansion which manifests that the nature of the interactions has a direct impact on the species abundance distribution. The central results are all species coexistence in a realistic range of the model parameters and that a certain discrete nature of the interactions induces multiple peaks in the species abundance distribution, providing the possibility of theoretically explaining multiple peaks observed in various field studies. To get more quantitative information, we also construct a non-perturbative theory which becomes exact on tree-like networks if all the species coexist, providing exact critical values of $u$ below which extinct species emerge. Numerical simulations in various different situations are conducted and they clarify the robustness of the presented mechanism of all species coexistence and multiple peaks in the species abundance distributions.
\end{abstract}

\maketitle

\section{Introduction}
The currently progressing serious reductions in diversity of ecosystems force us to consider in depth the relationships between ecosystem stability and species abundance distributions (SADs) quantifying community diversity~\cite{McCann:00,May:99}. Theoretical knowledge about SADs for systems at a given trophic level has greatly advanced in the last decade, based on Hubbell's neutral theory~\cite{Hubbell:01,Volkov:03,Alonso:06,Etienne:07}.
Meanwhile, two pioneering works~\cite{Gardner:70,May:72} opened the possibility of theoretical treatment of more complicated systems with multiple trophic levels, using linear models which continue to provide useful suggestions~\cite{Allesina:12}. 

Beyond the linear model, a nonlinear model called replicator dynamics (RD) has been employed to study ecosystems, and has offered qualitative knowledge about the global behavior of population dynamics~\cite{Rieger:89,Diederich:89,Oliveira:00,Oliveira:02,Oliveira:03,Tokita:04,Galla:06}. In the RD, the population of species evolves by its own fitness function which consists of two contributions: interactions with other species and self productivity of the species itself. Depending on the complicacy of the interactions, the RD yields various SADs~\cite{Oliveira:00,Oliveira:02,Oliveira:03,Tokita:04}. Although it is a highly nontrivial task to identify the fitness functions of species in a given real community, the RD provides a useful description of real ecosystems in a qualitative level in SADs. However, analytical treatment of it has so far been limited to the case where each species interacts with all other species. These “dense” interactions are not only seemingly unrealistic but also involve an undesirable simplification in SADs, due to the extensive sum of contributions in the fitness function. Thus, it is expected that novel and various SADs can be observed when “sparse” interactions, or specialist-specialist interactions, are employed.

\section{Problem setting}
Consider a community consisting of $N$ species, denote the fitness function of $i$th species by $F_i$, and assume that $F_i$ consists of two terms of pairwise interactions $J_{ij}$ and of self productivity $u$ 
\be
F_{i}(\V{x}|\V{J},u)=\frac{1}{2}\sum_{j(\neq i)}J_{ij}x_j-\frac{1}{2}u x_i,
\ee 
where $x_i(\geq 0)$ denotes the $i$th species' population. We assume the productivity $u$ is common among all the species, to purely see the effect of interactions. In the RD, each species is governed by the following equation 
\be
\frac{d x_{i}}{dt}=x_i \lb F_{i} -\bar{F}\rb,
\Leq{RE}
\ee
where the averaged fitness is introduced as 
\be
\bar{F}(\V{x}|\V{J},u)=\frac{\sum_{i}x_iF_{i}(\V{x}|\V{J},u)}{\sum_i x_i}.
\ee
The total population $\sum_{i=1}^{N}x_i$ is preserved in the RD, and without loss of generality we normalize this to the number of species as $\sum_{i=1}^{N}x_i=N$.
Properties of interactions crucially influence the dynamics. Here we treat “sparse” interactions: the number of interacting species of the $i$th species, $c_i$, is bounded by a fixed constant $c_{\rm max}$ independent of $N$. In addition, we investigate the case of symmetric relations, where $J_{ij} = J_{ji}$. This case is relatively simple, since the dynamics necessarily converges to a fixed point depending on the initial condition. Despite this simplicity, symmetric RDs can describe several communities such as a certain class of selection equation in population genetics and a classical model of community competing for resources~\cite{MacArthur:67}.

For symmetric interactions, the averaged fitness $\bar{F}$ plays the role of a Lyapunov function whose local maxima correspond to fixed points of the dynamics. Among those maxima, we focus on the global maximum stated as 
\be
\V{x}^*=
\argmax_{ \{ x_i \geq 0\}_{i=1}^{N} }
\lbb
\bar{F} \lb \V{x}| \V{J},u \rb
\rbb,
\Leq{extremization}
\ee
and the SAD is defined as
\be
P(x)=\frac{1}{N}\sum_{i=1}^{N}\delta\lb x-x_i^*\rb,
\ee
where $\delta(x)$ is the Dirac delta function. A typical shape of the SAD resembles a skewed lognormal distribution~\cite{Preston:62a,May:75}. We investigate how these quantities change if the interactions become sparse. 

\section{Result}
Our analysis is based on direct evaluation of \Req{extremization}, neglecting the non-negativity of $x_i$. This treatment provides exact results as long as no extinct species exist, which is the case for moderate values of $u$, but becomes less  precise as the proportion of extinct species grows. Despite this limitation, our analysis is sufficient to observe novel interesting SADs as seen below. 

Taking a direct variation of the averaged fitness with respect to $\V{x}$ on the above assumption and imposing the constraint $\sum_{i}x_i=N$, we get
\be
x_{i}^*=N\frac{\sum_{j}K^{-1}_{ij}}{\sum_{i,j}K^{-1}_{ij}},
\Leq{x-general}
\ee
where $K=uI-J$ and $I$ is the unit matrix. To obtain clear information from \Req{x-general}, the perturbative expansion with respect to $u^{-1}$ is performed under the assumption that $u$ is sufficiently large. The inverse of the matrix $K$ is expanded as
\be
K^{-1}=\lb u I - J \rb^{-1}=\frac{1}{u}\sum_{p=0}^{\infty} u^{-p}J^{p}.
\ee
Inserting this into \Req{x-general} gives
\begin{widetext}
\be
x^{*}_{i}=\frac{
1+u^{-1}\sum_{j}J_{ij}+u^{-2}\sum_{j,k}J_{ij}J_{jk}+u^{-3}\sum_{j,k,l}J_{ij}J_{jk}J_{kl}+\cdots
}{
1+u^{-1}\frac{1}{N}\sum_{i,j}J_{ij}+u^{-2}\frac{1}{N}\sum_{i,j,k}J_{ij}J_{jk}+
u^{-3}\frac{1}{N}\sum_{i,j,k,l}J_{ij}J_{jk}J_{kl}+\cdots
}.
\Leq{x-series}
\ee
\end{widetext}
Retaining higher order terms gives more precise results, but the computation becomes more complicated. Up to the second order of $u^{-1}$, the topology of the network does not affect the result and a clear discussion is possible. It is instructive to see an explicit formula of the SAD, the distribution of $x^{*}_i$, in the first-order expansion
\begin{widetext}
\be
P(x)\approx \int d\V{J} P(\V{J}) \delta\lb x- \lbb 1+u^{-1}\lb \sum_{j}J_{ij}-(1/N)\sum_{i,j}J_{ij} \rb   \rbb \rb,
\Leq{P(x)-first}
\ee
\end{widetext}
where $P(\V{J})$ is the distribution of the interactions $\V{J}$. Hence, if the distribution of interactions exhibits a certain discrete nature, i.e. if the marginal distribution $P_{ij}(J_{ij})\equiv \int \prod_{ \Ave{k,l}( \neq \Ave{i,j})  }  dJ_{kl} P(\V{J})$ consists of discriminable multiple peaks, the SAD correspondingly takes multiple peaks. The mechanism of multiple peaks here is based only on two assumptions: that $u$ is sufficiently large and that the distribution of interactions is discrete, providing the possibility of theoretically explaining multiple peaks observed in several field studies~\cite{Dornelas:08,Gray:05,Magurran:03}. Note that similar multiple peaks were observed in the RD with a specific distribution of dense interactions~\cite{Oliveira:02,Oliveira:03}, but we stress that our mechanism producing multiple peaks is completely different from the one in~\cite{Oliveira:02,Oliveira:03}. Their model's interactions are given by Hebb's rule of $p$ binary traits, meaning that all species are strictly categorized into $p+1$ groups by the symmetry in the $N\to \infty$ limit, and accordingly the SAD consists of $p+1$ (or less if multiple groups are extinct) delta peaks. In contrast to those delta peaks, our theory can naturally provide rounded discrete peaks and as well standard (lognormal like) functional forms by changing a single parameter $u$. Thus far such a flexibility has been absent in the densely interacting networks~\cite{Rieger:89,Diederich:89,Oliveira:00,Oliveira:02,Oliveira:03,Tokita:04,Galla:06}

One interesting outcome of treating sparse interactions is all species coexistence at moderate values of $u$. This coexistence starts to collapse at a critical value of $u_c$. A general upper bound of the critical value, $u_{c}^{\rm (UB)}$, can be derived by examining the condition where the numerator of \Req{x-series} vanishes. Each term in the numerator is bounded as
\be
\left| \sum_{j_1,j_2,\cdots,j_{p}}J_{j_1j_2}J_{j_2 j_3}\cdots J_{j_{p-1}j_p} \right| \leq c_{\rm max}^{p}J^{p}_{\rm \max},
\ee
where $J_{\rm max}$ is the maximum absolute value of the pairwise interaction $J_{ij}$. Thus, the numerator can be bounded from below
\be
&&
1+u^{-1}\sum_{j}J_{ij}+u^{-2}\sum_{j,k}J_{ij}J_{jk}+\cdots
\no \\
&& 
\geq 1-\sum_{p=1}^{\infty}\lb \frac{c_{\rm max} J_{\rm max}}{u} \rb^p=\frac{u-2c_{\rm max} J_{\rm max}}{u-c_{\rm max}}.
\Leq{u_c-bound}
\ee
This gives the general upper bound as 
\be
u_c^{\rm (UB)}=2c_{\rm max}J_{\rm max}.
\Leq{upper bound}
\ee
As long as $u>u_c^{\rm (UB)}$, all species coexistence is guaranteed. This is actually observed in \Rfig{PD} for a specific choice of $P(\V{J})$ (see below for the detail). A similar general upper bound can be derived even if the self interactions can vary depending on species as long as its mean is large enough.  In that case, the expansion with respect to $J/u$ is replaced to the one with respect to $J/U$ where $U$ is a diagonal matrix whose entries are species-dependent self interactions. 

Beyond the perturbation and the general bounding method, we can provide a non-perturbative theory. Terms of higher orders than $2$ can be categorized into those with and without loops. For example, in the case of third-order terms, $\sum_{j,k,l}J_{ij}J_{jk}J_{kl}$, terms of $J_{ij}J_{jk}J_{ki}~(j\neq i,j\neq k,k\neq i)$ have a closed loop, and other terms do not have loops. It is known that summing terms without loops is possible for all  orders in a systematic manner, which is known to be equivalent to the so-called Bethe approximation, or cavity method~\cite{ADVA,INFO,Kabashima:12}. For a single instance of network, this method evaluates influences to a newcomer species from an existing species, which are termed cavity biases, in a self-consistent manner. 

For clarity of the following discussion based on the cavity method, we treat a single degree network $c_i=c$, and fix the distribution of interactions as 
\be
P(J_{ij})=\frac{1+\Delta}{2}\delta(J_{ij}-1)+\frac{1-\Delta}{2}\delta(J_{ij}+1).
\Leq{P(J_{ij})}
\ee
This simple function maintains a discrete nature and also has a parameter, $\Delta$, controlling the ratio of mutualistic and competitive relations. This setup enables us to see the effects of productivity $u$, of the ratio of mutualistic relations $\Delta$, and of the degree of network $c$, in an unified manner. The distribution of the cavity biases, $\hat{q}(\hat{H})$, satisfies the following self-consistent equation:
\be
\hspace{-5mm}
\hat{q}(\hat{H})=\int\prod_{l=1}^{c-1}d\hat{H}_{l}\hat{q}(\hat{H}_{l})
\overline{
\delta \lb \hat{H}-J\hat{a} \lb r+\sum_{l=1}^{c-1}\hat{H}_{l}  \rb   \rb
},
\Leq{B2B}
\ee
where $\overline{\cdots}$ denotes the average over the interaction $J$, and $\hat{a}$ and $r$ are given by the external parameters as
\be
&&
\hat{a}=\frac{u - \sqrt{u^2-4(c-1)}}{2(c-1)},
\\
&&
r=( u-c\hat{a} )\lb 1- \frac{ c\hat{a}\Delta }{ 1+\hat{a}\Delta } \rb.
\ee
Using the solution of \Req{B2B}, the SAD $P(x)$ is assessed as
\be
P(x)=
\int \prod_{l=1}^{c} d\hat{H}_l \hat{q}(\hat{H}_l) \, \delta\lb x- \frac{r+\sum_{l=1}^{c}\hat{H}_l}{u-c\hat{a}}\rb.
\Leq{P(x)}
\ee
Solving \Req{B2B} and inserting the solution into \Req{P(x)} give the results shown in \Rfigss{PD}{diversity-u}. This gives an exact treatment for the interaction network without loops, as well as for the randomly generated network which has some global loops but their influence can be neglected in the large size limit.

\begin{figure}[t]
\begin{center}
  \includegraphics[width=0.7\columnwidth]{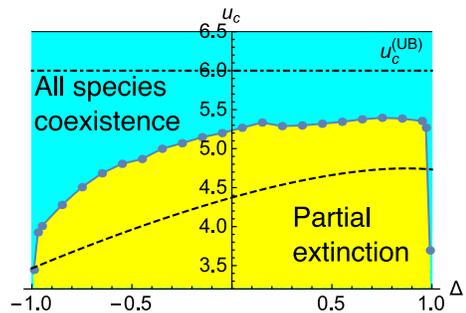}
  \caption{Critical values of productivity $u_c$ plotted against $\Delta$ for $c=3$. The points are the non-perturbative results, and the dashed line corresponds to the second-order perturbation. Both of the values locate well below the general upper bound $u_{c}^{\rm (UB)}=2c=6$. }
\Lfig{PD}
\end{center}
\end{figure}
\Rfig{PD} displays $u_c$ plotted against $\Delta$ for $c=3$. The points are based on the non-perturbative theory, and the dashed line is derived by keeping only up to the second-order terms of $u^{-1}$ in \Req{x-series}. The qualitative shape of the $u_c$ curve is captured by the second-order approximation, but the quantitative deviation is not small.  An interesting, and perhaps somewhat counter-intuitive, observation is the dependence on $\Delta$. Larger values of $\Delta$ provide more mutualistic relations, and hence \Rfig{PD} shows more mutualistic communities tend to be more extinct-prone, in the sense that extinct species start to appear even at larger $u$. The critical value $u_c$ drastically drops off around $\Delta=\pm1$, which exhibits singularities at those points. This is natural since $\Delta=\pm 1$ corresponds to the case where all the species are equal and thus $x_i=1(\forall i)$ holds irrespective of the value of $u$. We also calculate the $c$-dependence of $u_c$ for fixed $\Delta$ and find that $u_c$ monotonically increases as $c$ grows.

\begin{figure*}
\begin{center}
  \includegraphics[width=0.6\columnwidth]{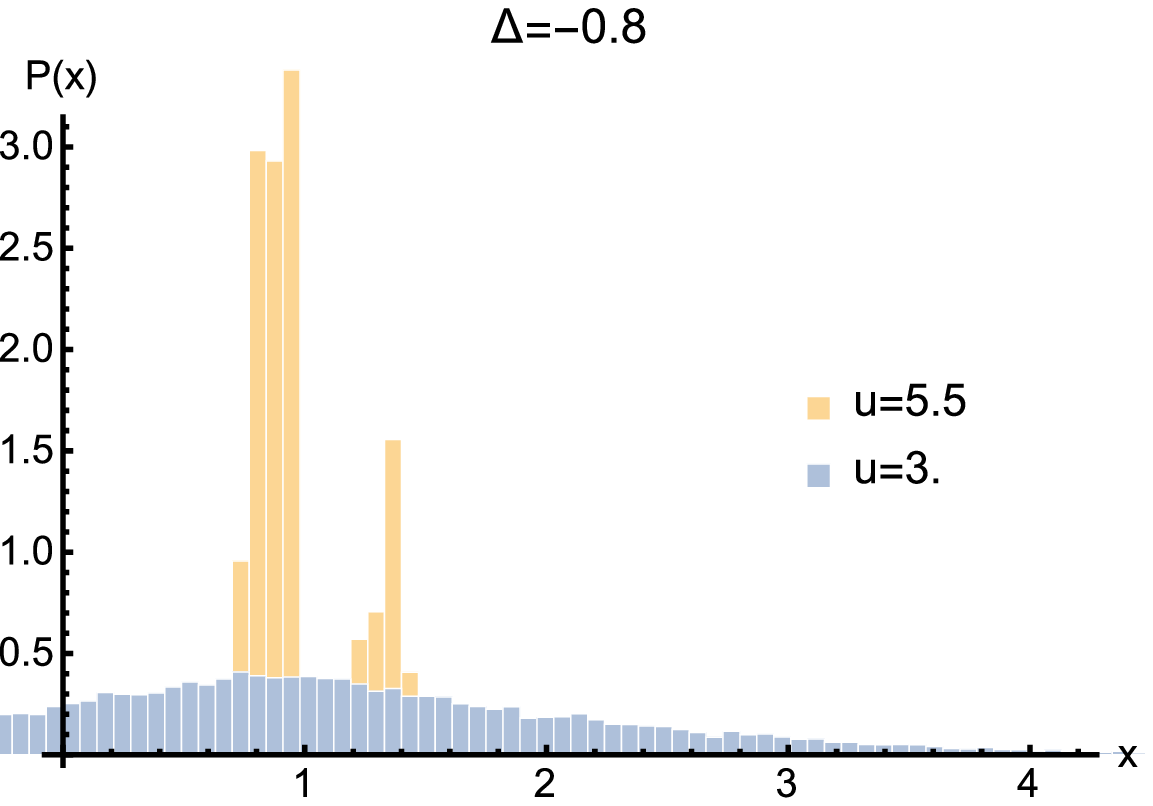}
  \includegraphics[width=0.6\columnwidth]{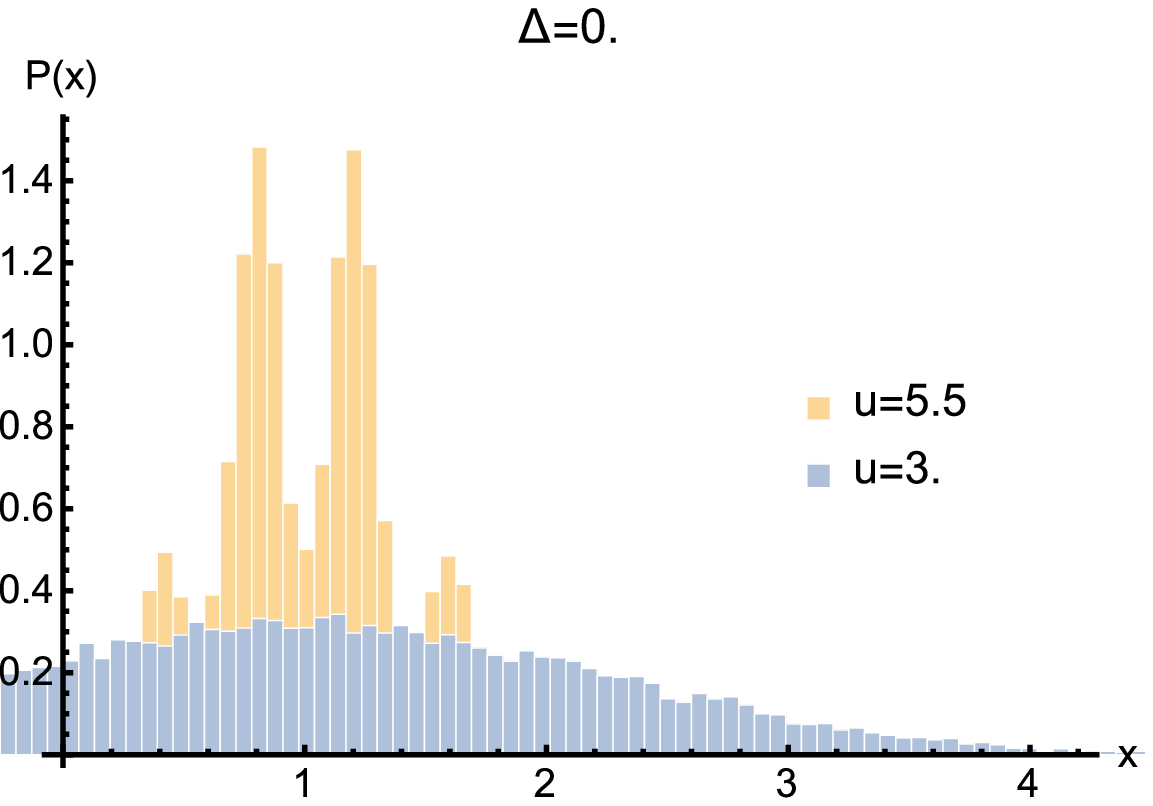}
  \includegraphics[width=0.6\columnwidth]{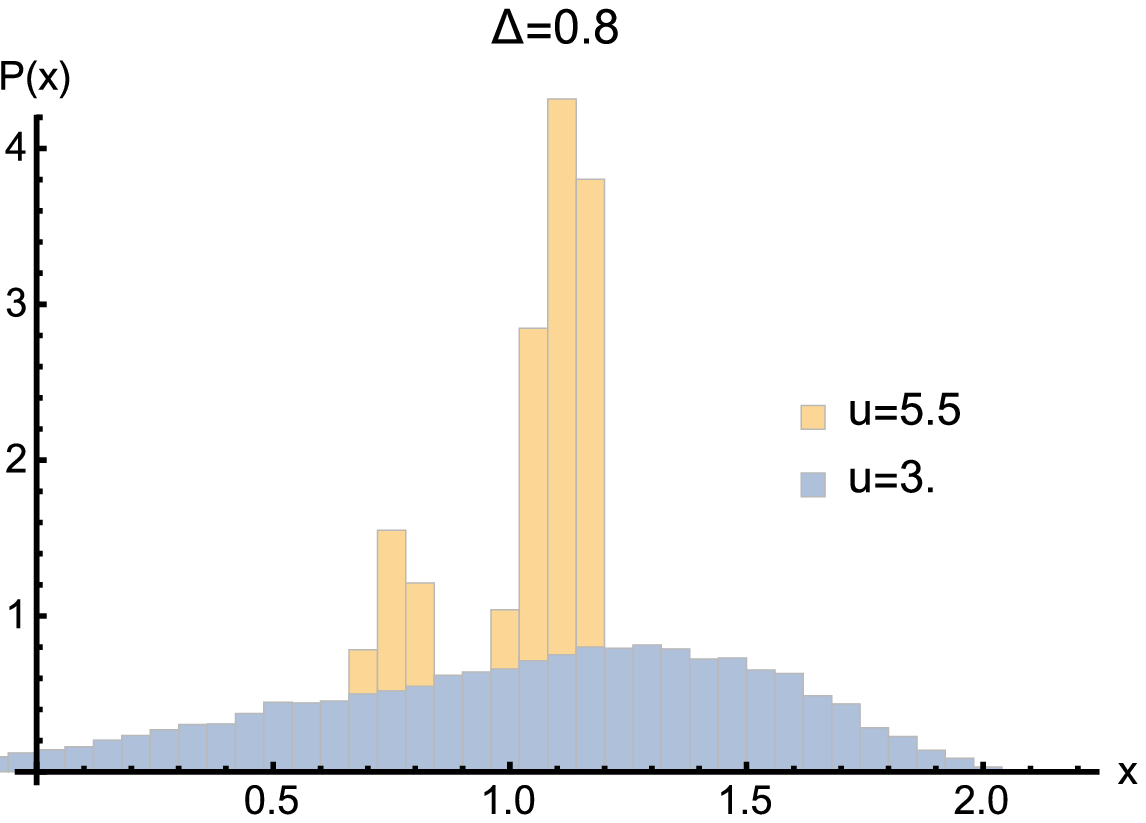}
\caption{ The SADs for $\Delta=-0.8,0$ and $0.8$ at $c=3$.}
\Lfig{SADs}
\end{center}
\end{figure*}
\Rfig{SADs} shows the SAD for several values of $u$ and $\Delta$ at $c=3$. For $u=5.5(>u_c)$, the distribution is symmetric about $x=1$ for $\Delta=0$, though it is biased to $x>1$ or to $x<1$ for $\Delta \neq 0$. For the competitive case $\Delta=-0.8$, the largest peak appears with $x<1$, and the long tail persists in the $x>1$ region, while for the mutualistic case $\Delta=0.8$ the opposite is true.  As $u$ decreases, the discreteness becomes weaker and the functional forms become closer to a standard skewed lognormal distribution, irrespective of $\Delta$, as seen at $u=3(<u_c)$. Hence, our theory smoothly connects the standard functional form and that with multiple peaks, by a single parameter $u$. The $c$-dependence of the SAD appears as the number of peaks for large $u$ which is equal to $c+1$, as understood by \Req{P(x)-first}. For small $u$, tails of the SADs in $x<0$ appear because we neglect the non-negativity constraint. These tails mean there occurs a partial extinction in the corresponding parameters. In this situation, our analysis is not exact, but it is possible to interpret the cumulative distribution in $x<0$ as an approximation of the ratio of the extinct species.

\begin{figure}
\begin{center}
  \includegraphics[width=0.48\columnwidth]{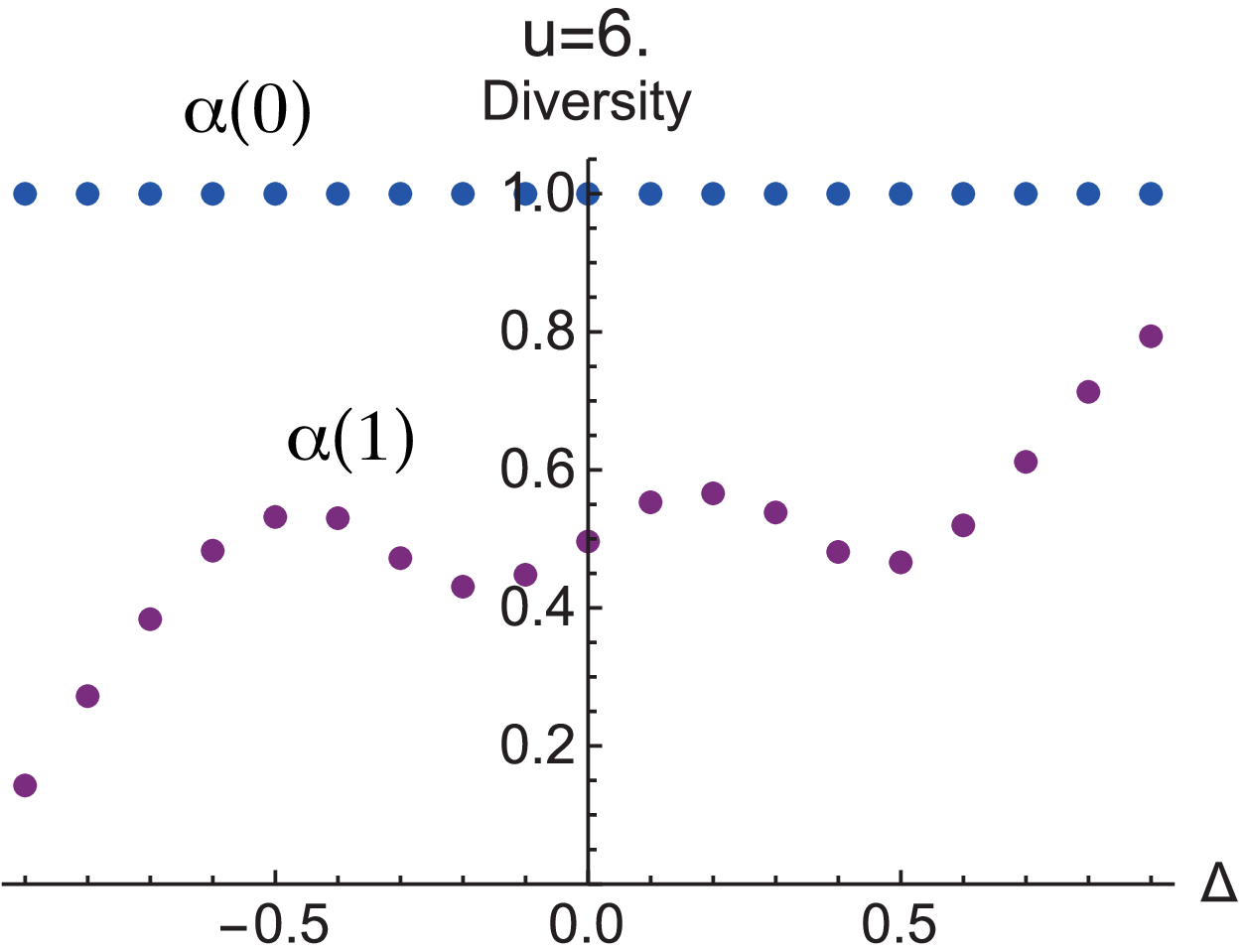}
  \includegraphics[width=0.48\columnwidth]{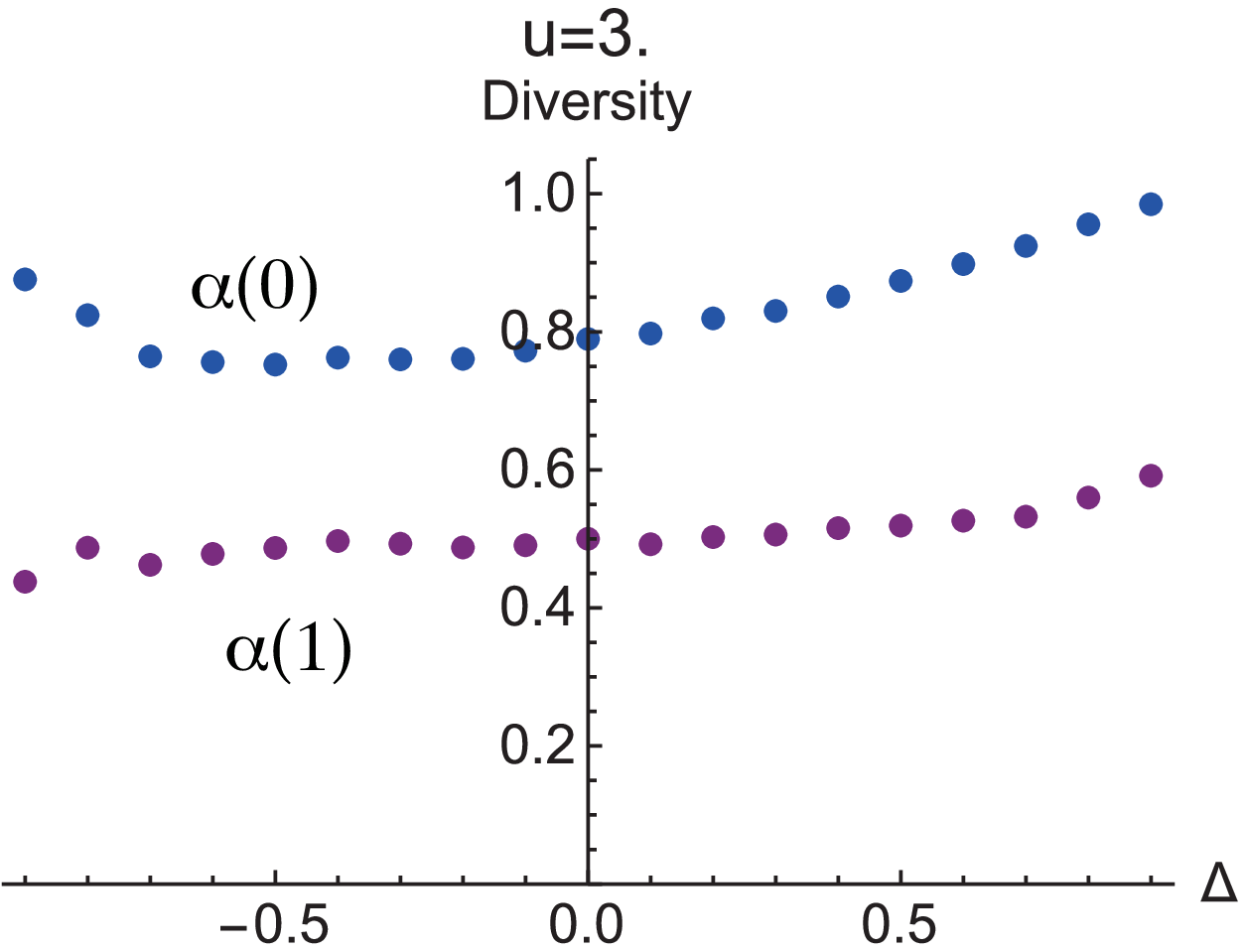}
\caption{The diversity $\alpha(0)$ and $\alpha(1)$ are plotted against $\Delta$ for $u=6$ (left) and $3$ (right) at $c=3$. The dependence on $\Delta$ is not monotonic. }
\Lfig{diversity-u}
\end{center}
\end{figure}
An interesting finding appears in the survival function $\alpha(y)=1-\int_{0_{-}}^{y}P(x)dx$, an index quantifying community diversity~\cite{Tokita:04}. \Rfig{diversity-u} shows $\alpha(1)$ and $\alpha(0)$ plotted against $\Delta$ for $u=6(>u_c)$ and $u=3(<u_c)$ at $c=3$. We observe non-monotonic behavior of diversity, 
and particularly $\alpha(1)$ shows an oscillating behavior as $\Delta$ changes. This is related to the multiple peaks of the SAD: the height and location of the highest peak and the tail sensitively change as $\Delta$ deviates, leading to nontrivial behavior of the diversity. 

To see the robustness of SAD discreteness in large $u$ regions discussed so far, we perform a numerical simulation. \Rfig{hist-univ} shows the result for the case where the interaction network is a random graph of the degree $c=3$ and the values of $J_{ij}$ are drawn from a sum of two Gaussian distributions 
\be
P(J_{ij})=\frac{1}{\sqrt{8V\pi}}\lbb 
e^{-\frac{(J_{ij}-1)^2}{2V} }+e^{ -\frac{(J_{ij}+1)^2}{2V} }
\rbb,
\Leq{J-dist-Gauss}
\ee
with $V=0.1$. Two peaks are symmetrical and thus correspond to $\Delta=0$. The result clearly demonstrates the presence of multiple peaks. This is not trivial, since the value of $J_{ij}$ drawn from \Req{J-dist-Gauss} is not bounded. 
\begin{figure}
\begin{center}
  \includegraphics[width=0.7\columnwidth]{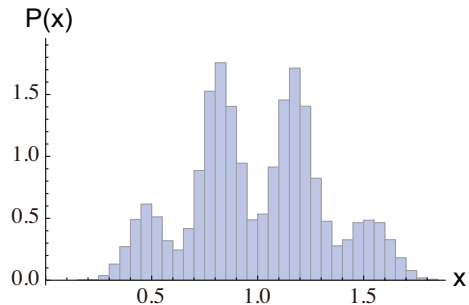}
\caption{The SAD with the interactions generated from \Req{J-dist-Gauss}, the network structure of which is a random graph of single degree $c=3$. The parameters are $u=6.03$, $N=16000$. }
\Lfig{hist-univ}
\end{center}
\end{figure}
We also perform other numerical simulations in several different situations, and the result is given in \Rfig{SAD-several}. The left panel is for the square lattice; $N=16384$ and $u=8$. The middle panel is for a heterogeneous random network with degrees $c=3,4,5,6$, each ratio of which is $p=0.45,0.35,0.1,0.1$, respectively; $N=1000$ and $u=10$. The right panel is for random self interactions, whose values are uniformly taken from $( 9.4,10.6 )$, and the network is single degree $c=3$ and the simulated size is $N=1000$. In all these cases the interactions are unbiased binary. 
\begin{figure*}
\begin{center}
\vspace{0mm}
\includegraphics[width=0.6\columnwidth]{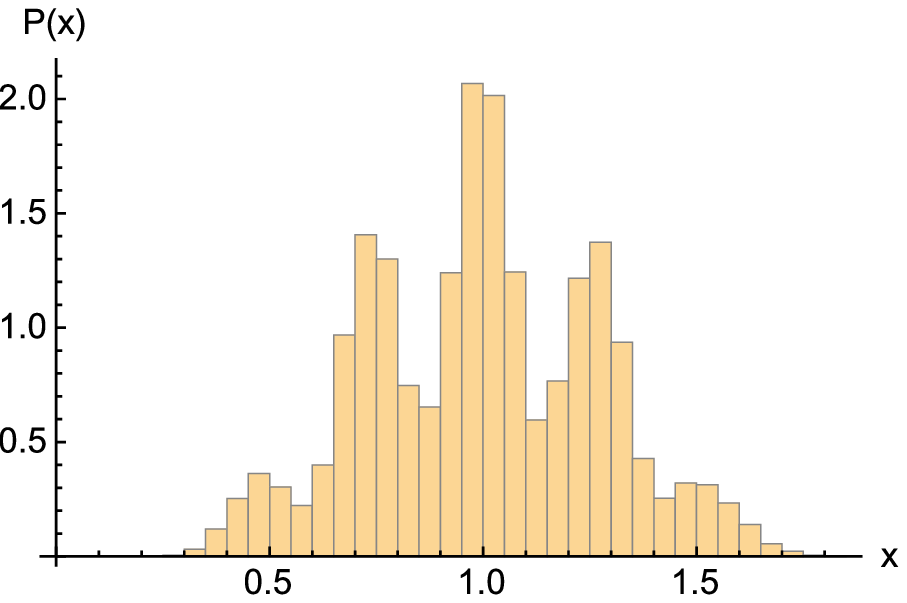}
\includegraphics[width=0.6\columnwidth]{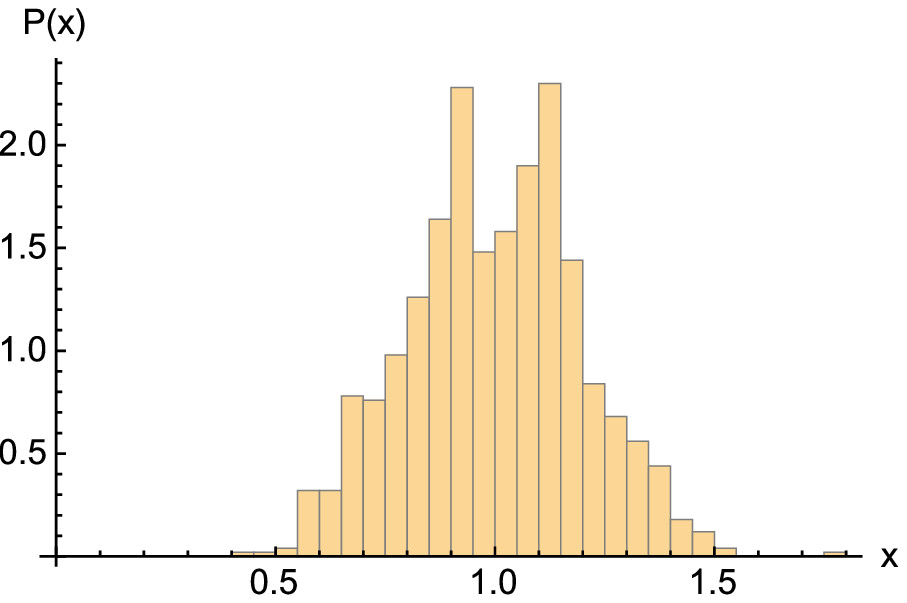}
\includegraphics[width=0.6\columnwidth]{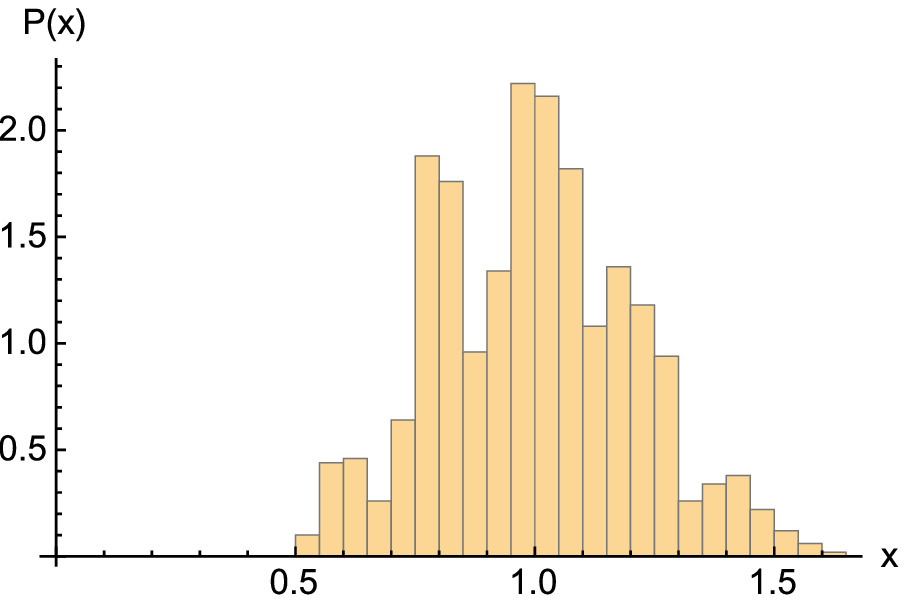}
\vspace{0mm}
\caption{The SADs for the square lattice with unbiased binary interactions (left), a heterogeneous random network with degrees $c=3,4,5,6$ and unbiased binary interactions (middle), and random self interactions on a single degree network $c=3$ with unbiased binary interactions. }
\Lfig{SAD-several}
\end{center}
\end{figure*}

All these numerical simulations universally exhibit multiple peaks in the SADs, as long as $P(J_{ij})$ has a discrete nature and the self interactions are sufficiently large, and thus the presented mechanism is fairly robust. Therefore, we again stress the importance of this mechanism in explaining actually observed multiple peaks in several field studies~\cite{Dornelas:08,Gray:05,Magurran:03}.

\section{Conclusion}
Our analysis of the RD with sparse interactions has provided several nontrivial behavior in the SADs: coexistence of all the species, multiple peaks. These consequences were transparently understood from the perturbative expansion assuming the self interaction $u$ is large enough and the interactions have a certain discrete nature.  The general upper bound of critical value of $u$, below which extinct species emerge, has also been evaluated.

We also provided a non-perturbative theory which is exact on tree-like networks without loops, to obtain more quantitative information. As a result, exact critical values of $u$ and the SADs' dependence on model parameters have been calculated for a specific network.  We have found a nontrivial dependence of diversity on the ratio of mutualistic relations and a drastic change of the abundance distribution's shape from the one with multiple peaks to a standard skewed lognormal distribution. We stress that this drastic change is controlled by a few parameters, the self interaction $u$ and the ratio of mutualistic relations $\Delta$. Thus our theory provides a possibility of unifying different shapes of the abundance distributions. Comparison with experimental observations is highly desired. 

Our results so far are derived by investigating the RD with a few assumptions, and thus they can be applied to other contexts in which the RD appears, such as population genetics, game theory, and chemical networks in living cells~\cite{Hofbauer:98,Nowak:06}.

\section*{Acknowledgements} This work was supported by a Grant-in-Aid for JSPS Fellows (No. 2011) (TO),  KAKENHI Nos. 26870185 (TO), 25120013 (YK), and 24570099 (KT). KT also acknowledges support in part by the project ``Creation and Sustainable Governance of New Commons through Formation of Integrated Local Environmental Knowledge'', at the Research Institute for Humanity and Nature (RIHN), and the project ``General Communication Studies'', at the International Institute for Advanced Studies (IIAS).


\end{document}